\documentclass[twocolumn,superscriptaddress,showpacs,amsfonts]{revtex4-1}
\pdfoutput=1

\usepackage{epsfig}
\usepackage{graphicx}
\usepackage{epsfig}
\usepackage{graphicx,graphics}

\begin{document}
\title{Interface currents and magnetization in singlet-triplet superconducting heterostructures: Role of chiral and helical domains}

\author{Alfonso Romano}
\affiliation{CNR-SPIN, I-84084 Fisciano (Salerno), Italy and
Dipartimento di Fisica ``E. R. Caianiello'', Universit\`a di
Salerno, I-84084 Fisciano (Salerno), Italy}
%
\author{Canio Noce}
\affiliation{CNR-SPIN, I-84084 Fisciano (Salerno), Italy and
Dipartimento di Fisica ``E. R. Caianiello'', Universit\`a di
Salerno, I-84084 Fisciano (Salerno), Italy}
\author{Ilya Vekhter}
\affiliation{Department of Physics and Astronomy, Louisiana State
University, Baton Rouge, Louisiana, 70803, USA}
\author{Mario Cuoco}
\affiliation{CNR-SPIN, I-84084 Fisciano (Salerno), Italy and
Dipartimento di Fisica ``E. R. Caianiello'', Universit\`a di
Salerno, I-84084 Fisciano (Salerno), Italy}

\begin{abstract}
Chiral and helical domain walls are generic defects of topological spin-triplet superconductors. We study theoretically the magnetic and transport properties of superconducting singlet-triplet-singlet heterostructure as a function of the phase difference between the singlet leads in the presence of chiral and helical domains inside the spin-triplet region. The local inversion symmetry breaking at the singlet-triplet interface allows the emergence of a static phase-controlled magnetization, and generally yields both spin and charge currents flowing along the edges. The parity of the domain wall number affects the relative orientation of the interface moments and currents, while in some cases the domain walls themselves contribute to spin and charge transport. 
We demonstrate that singlet-triplet heterostructures are a generic prototype to generate and control non-dissipative spin and charge effects, putting them in a broader class of systems exhibiting spin-Hall, anomalous Hall effects and similar phenomena. Features of the electron transport and magnetic effects at the interfaces can be employed to assess the presence of domains in chiral/helical superconductors.
\end{abstract}

\maketitle

\section{Introduction}

In the past two decades the study of superconductor-based heterostructures has increasingly been a focal center both for their potential application in novel electronic devices and because of the richness of the underlying fundamental physics.
Even more than in the traditional heterostructures, for superconducting systems the interface controls the symmetry and the nature of the emerging electronic states, and hence sets the physical properties. Interface potentials and the associated electronic reconstruction lead to exotic proximity effects, edge states, and possible spontaneous symmetry breaking as well as the unusual spin and charge electronic transport.

Whether gapped systems, such as
insulators or superconductors, exhibit robust protected
low-energy states at their boundary depends on the symmetries of their
bulk electronic states~\cite{Hasan2010Rev,Qi2010Rev,Moore2010,d-wave,Eschrig:PT}.
The earliest and most prominent example of such topological state is the quantum Hall state identified by the topological number introduced
by Thouless, Kohmoto, Nightingale, and den Nijs
(TKNN)~\cite{Thouless1982}.
Among the superconducting systems, a notable case of superconductor with non-trivial TKNN number is the two dimensional chiral ($p$+i$p$)-wave superconductor with
time-reversal symmetry-breaking order parameter, and the leading candidate for its realization is
Sr$_2$RuO$_4$~\cite{Maeno98,Mackenzie2003,Kallin2012}.
Recent intense interest in this area led to the identification~\cite{Kane2005,Moore2007,Fu2006,Fu2007,Fukui2007,Sheng2006,Qi2006,
Schnyder2008,Roy2006,Tanaka2012} of additional classes of
topological superconductors with time reversal invariance.
These can be viewed as the time reversal partners
of the chiral ones, just as the quantum spin Hall systems relate to integer quantum Hall systems.
In contrast to the chiral superconductors and in analogy with quantum spin Hall systems, topological time-reversal-invariant superconductors
can have zero modes that come in pairs, due to Kramers's degeneracy, and can support
counter-propagating helical states of opposite spins near the boundary that carry a net spin current.
Among the
candidate materials, where this effect may occur, there are the $^3$He B phase~\cite{Voll90,Schnyder2008,Chung2009}, Cu-doped BiSe$_2$~\cite{Fu2010,Hsieh2012,Sasaki2011},
p-type TlBiTe$_2$~\cite{Yan2010}, the interface state of Sr$_2$RuO$_4$ ~\cite{Tada2009}, BC$_3$~\cite{Chen2014}, and even
doped Mott insulators~\cite{Hyart2012,You2012,Okamoto2013}.



A distinctive mark of these triplet phases is that the topological nature relies on the orbital degeneracy of the superconducting order (for example between $p_x$ and $p_y$ state), allowing for the existence of the domains ($p_x\pm ip_y$). Since the degeneracy of the most favorable superconducting state is discrete, the domain walls separating such regions
are well defined, and create spatial variations of the order parameter that give rise to subgap electronic states\cite{Mukherjee2015},  in close analogy to what happens at the surface.
In the example above, chiral $p$-wave superconductivity exhibits two-fold degeneracy corresponding to clockwise
or counterclockwise winding of the orbital superconducting phase for each spin orientation~\cite{Matsumoto1999},
allowing for two types of chiral domains separated by a chiral domain wall (chiral-DW) (see Fig. 1).
While up to now there has been no direct observation of the chiral-DW~\cite{Kirtley2007}, its existence has been strongly suggested by transport studies in Sr$_2$RuO$_4$ junctions~\cite{Kidwingira2006,Kambara2008,Anwar2013}.
Such a domain wall serves as a one-way channel for charge transport, with non zero conductance measured between a pair of metal contacts~\cite{Serban2009}, allowing probes of chiral superconductivity via electrical measurements.

In a similar fashion, in helical superconductors with time-reversal invariance a domain wall (helical-DW) can occur between regions with opposite orbital winding for each spin orientation of the Cooper pairs.  For instance, in non-centrosymmetric superconductors, where the crystal structure dictates the form of the triplet superconducting component, two regions with the opposite inversion symmetry breaking fields face each other across twin boundaries. In such systems with dominant odd-parity pairing twin boundaries may exhibit helical edge modes~\cite{Sato2009,Lu2010,Tanaka2009,Yokoyama2005}, akin to the electronic edge states of the quantum spin Hall insulator, and also provide possible realizations  for additional spontaneous symmetry breaking and anomalous vortices enclosing fractional fluxes~\cite{Iniotakis2008,Mukuda2009,Arahata2013}.


It is well established that the properties of the surface states in topological superconductors can be manipulated in heterojunctions with conventional superconducting materials. Both spin and charge currents, as well as the magnetic moments emerging at the interface, sensitively depend on the nature of the pairing interaction and the interface potentials~\cite{Romano2013}, and can be controlled by the phase difference across the junction~\cite{Romano2016}. At the same time, the effect of the domain walls on the properties of such junctions has not been previously explored.

The aim of the paper is therefore to investigate the role of domain walls inside the spin-triplet superconducting region for the generation and control of magnetization, spin and charge currents at the interface between chiral/helical spin-triplet $p$-wave superconductors with conventional
spin-singlet $s$-wave superconductors in the presence of a phase difference across the heterostructure.
The response of the resulting singlet-triplet-singlet (S-T-S) superconducting planar junction, schematically depicted in Fig. 1,
is analyzed for each of the two cases of time reversal symmetry (TRS) breaking chiral or TRS-preserving helical  spin-triplet order parameters, comparing
single-domain and two-domain spin-triplet layers.
Since the chiral (helical) spin triplet states have edge modes with spontaneously flowing charge (spin) currents, different types of
configurations can occur close to each interface, which are intimately connected to the possible occurrence of a non-vanishing spin polarization as due to the local inversion symmetry breaking at the singlet-triplet interface and the subsequent parity mixing.
The emerging physical scenario can be quite rich, with currents with variable directions and spin polarization with different orientations that can be tuned by the phase difference applied
between the singlet layers and that can combine to give a net charge or a net spin flow at the interface separating different domains (see Fig. 1).

Some of the features, such as static magnetic moments or certain components of the spin currents, develop solely at the interfaces between the singlet and the triplet states, and depend sensitively on the spin structure of the triplet order parameter. While their existence at each interface does not depend on the existence of the domain walls, their relative orientation does depend on the number of domain walls (parity-DW). Hence the domain walls directly control the net values of these quantities across the junction, and the phase difference across the heterostructure allows sensitive control of their magnitudes. In other cases, the domain wall directly contributes to the components of the spin and charge current, often dominating the contribution from the S-T boundaries. In these circumstances the phase-sensitivity is weaker, but the domain wall contribution is more pronounced. We explore these possibilities and give detailed analysis of the behavior of the magnetization, spin and charge currents in each configuration.  In principle, we envision the possibility of having a switchable functional heterostructure with distinct possible values of the integrated amplitude of the magnetization, spin- and charge- currents, either null (or very small) or substantial. In a very broad sense, these effects can be seen as counterparts of the spin-Hall, anomalous Hall, and their inverse in the superconducting state.
\begin{figure*}[t!]
\includegraphics[width=0.95\textwidth]{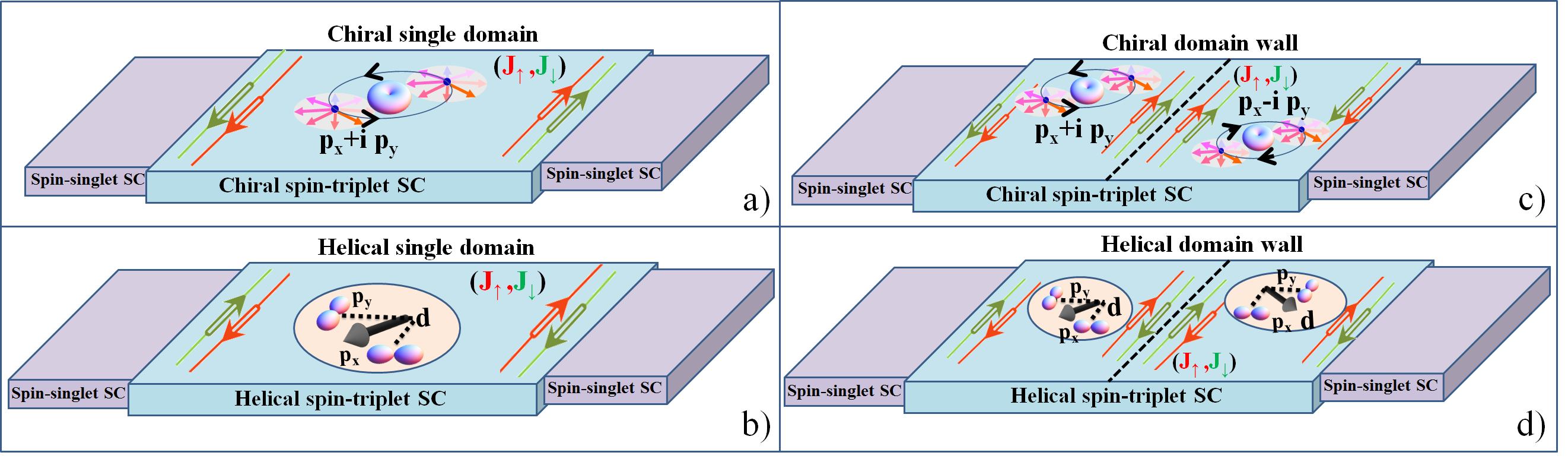}
\caption{(color online).
Schematic representation of the singlet-triplet-singlet (S-T-S) heterostructure in the presence of a spin-triplet superconductor with single (a) or double chiral domain (c) or single (b) or double (d) helical domain. In (c) and (d) the domain wall is indicated with a black dashed line to separate the two regions with opposite winding of the superconducting order parameter for each spin direction. The arrows (green and red for up and down spin) along the domain wall and at the singlet-triplet interface indicate the spin dependent charge currents due to the presence of Andreev bound states.}
\label{fig:1}
\end{figure*}
%

%
%

%
%

\section{Model and methodology}

We consider a planar S-T-S trilayer of size $L\times L$ (in units of the lattice constant) extending in the $x$-$y$ plane, where the two interfaces separating the central $p$-wave spin-triplet superconductor from the two lateral $s$-wave spin-singlet ones are taken to be parallel to the $y$ direction.  For simplicity we choose the layers of equal width, so that, if we denote the lattice sites by $\mathbf{i}\equiv(i_x,i_y)$, with $i_x$ and $i_y$  integers between $-L/2$ to $L/2$, the two singlet-triplet interfaces are located at $i_x=\pm L/6$. Asymmetry of the junction geometry does not qualitatively influence the physical behavior of the heterostructure, and does not affect our conclusions.  When a spin-triplet layer made of two domains with opposite chirality/helicity is considered, the sites ($i_x$=0,$\,i_y$) define the boundary separating the two regions.

The Hamiltonian is defined as
\begin{equation}
H=H_0+H_S+H_T
\label{eq:H}
\end{equation}
with
\begin{eqnarray}
H_0 &=& \sum_{\langle \mathbf{i} ,\mathbf{j} \rangle \in S,\,\sigma}
t_{\mathbf{i} ,\mathbf{j}} (c^{\dagger}_{\mathbf{i}\,\sigma}
c_{\mathbf{j}\,\sigma}+\text{h.c.}) -\mu \sum_{\mathbf{i} \in S,\sigma}
n_{\mathbf{i}\sigma} \nonumber \\
H_S& =& \sum_{\mathbf{i}\in S} U_0 n_{\mathbf{i} \uparrow} n_{\mathbf{i}\downarrow}
\nonumber \\
H_T&=& \sum_{\langle \mathbf{i} ,\mathbf{j} \rangle \in T} V_{\uparrow \downarrow}\left( n_{\mathbf{i} \uparrow}
n_{\mathbf{j}\downarrow}+n_{\mathbf{i}\downarrow} n_{\mathbf{j}
\uparrow} \right) - \sum_{\langle \mathbf{i} ,\mathbf{j} \rangle , \sigma} V_{\sigma \sigma} \, n_{\mathbf{i} \sigma}
n_{\mathbf{j}\sigma} \;. \nonumber \\
\end{eqnarray}
where $H_0$ contains the single particle terms, while $H_S$ and $H_T $ describe the pairing in the spin-singlet and triplet regions of the junction, respectively.
Here, $c_{\mathbf{i}\,\sigma}$ is the annihilation operator of an electron with spin $\sigma$ at the site ${\mathbf{i}}$, $n_{\mathbf{i}\,\sigma} = c^{\dagger}_{\mathbf{i}\,\sigma} c_{\mathbf{i}\,\sigma}$ is the spin-$\sigma$ number
operator and $t_{\mathbf{i}\mathbf{j}}$ is the hopping amplitude that is nonvanishing only between the nearest neighboring
sites $\langle \mathbf{i},\mathbf{j}\rangle $, with $\mu$ being the chemical potential. Periodic boundary conditions are assumed only in the $y$ direction, since the presence of the interfaces breaks the translational symmetry along $x$.
The short ranged (i.e. nearest-neighbor attractive interaction) $-V_{\uparrow\downarrow}$ $(V_{\uparrow\downarrow}>0)$ allows for
both singlet and triplet pairing channels with zero spin projection along the $z$ axis, whereas $-V_{\sigma \sigma}$ $(V_{\sigma \sigma}>0)$ is effective only for the equal-spin triplet channel. $-U_0$ is the superconducting coupling for the local $s$-wave spin-singlet configuration in the lateral sides of the junction. Since we deal with magnetic effects at the singlet-triplet interface, it is convenient to introduce the local spin density polarization $\vec{s}(\mathbf{i}) =\sum_{s,s'}c^{\dagger}_{\mathbf{i}\,s}{\vec{\sigma}}_{s,s'}c_{\mathbf{i}\,s'}$ and the averaged quantities $\vec{S}(i_x) =\frac{1}{Ly} \sum_{i_y} \sum_{s,s'} \langle c^{\dagger}_{i_x i_y\,s}{\vec{\sigma}}_{s,s'}c_{i_x i_y\,s'} \rangle$ for the total magnetization at a given position along the $x$-direction, $i_x$. The total magnetization for any part of the S-T-S heterostructure is simply the sum over the sites $i_x$.  For our purposes, it is useful to consider two distinct ranges for the computed integrated quantities. Below we determine the expectation values of the magnetization for the whole system by summing up over all the sites $i_x$ in $[-L/2,L/2]$, and also for half of the heterostructure within the interval $[-L/2,0]$. The latter is needed to make connections with the results for a single interface studied previously.

The edge states at the singlet-triplet boundary can support net currents
whose spin character depends on the nature of the chiral and helical triplet state, the mixed-parity configuration emerging at the interface, and the occurrence of domains within the spin-triplet superconductor. Study of the spin and charge currents in the S-T-S heterostructure are one of the foci of our attention below. The local values of the current with spin component $\alpha$ flowing along the S-T interface at the site $i_x$ is
\begin{equation}
J^{\alpha}_s (i_x)= \frac{2\,t}{L_y} \sum_{p_y} \sin(p_y) \langle c^{\dagger}_{i_x p_y \nu} \sigma^{\alpha}_{\nu\nu^\prime} c_{i_x p_y \nu^\prime} \rangle \,
\label{eq.J}
\end{equation}
where $c^{\dagger}_{i_x p_y \nu}$ is the creation operator of an electron at the site $i_x$ with a given momentum $p_y$ along the interface, obtained by performing the Fourier transformation only for the $i_y$ coordinates. $\sigma^\alpha$ is the Pauli matrix corresponding to the $\alpha$ spin direction.  The charge current, $J_c(i_x)$, is then obtained by summing the contribution of the up and down polarized electrons, i.e. by replacing the Pauli matrix with the identity matrix above. Similar to the magnetization, we evaluate average quantities that include the summation of all the currents at different distances from one of the two singlet-triplet interfaces, i.e. $J^{z}_{s}=\frac{1}{L_y} \sum_{-L/2<i_x<0} J^{z}_{s} (i_x)$.

\begin{figure*}[t!]
\includegraphics[width=0.95\textwidth]{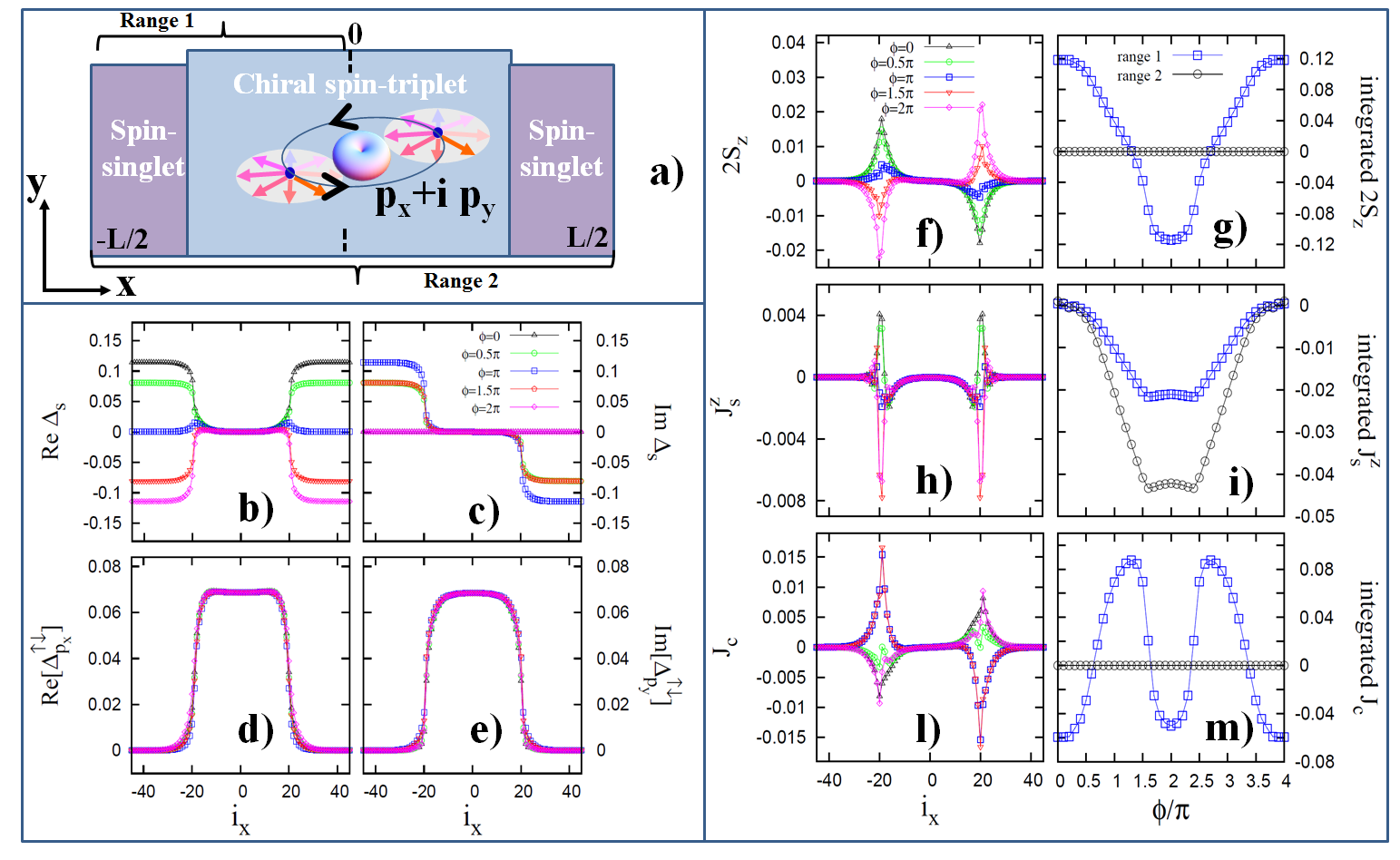}
\caption{(color online). (a) schematic description of the singlet-triplet-singlet (S-T-S) heterostructure with a single chiral domain in the spin-triplet region having $p_x+i p_y$ orbital symmetry and zero spin projection along the $z$ direction. The encircling arrows for the electron pair are used to sketch the coherent spin-triplet state with zero projection along the direction perpendicular to the spin plane.  The donut-like shape schematically indicates the electron spatial probability associated to the $p_x+i p_y$ orbital state. Range 1 (2) indicates half (entire) extension of the S-T-S system. Spatial profile of the real ((b) and (d)) and imaginary ((c) and (e)) parts of spin-singlet and spin-triplet order parameters, respectively, is shown at different values of the phase difference $\phi$ between the spin-singlet sides of the heterostructure. Panels (f), (h) and (l) indicate the spatial evolution of the $z$-projected magnetization (i.e. 2$S_Z$), $z$-component of the spin-current ($J^{Z}_{S}$) and charge-current ($J_C$), respectively. (g), (i) and (m) describe the phase dependent behavior of the integrated quantities over the range 1 (blue squares) and 2 (black circles) for $S_Z$, $J^{Z}_{S}$, and $J_C$, respectively.}
\label{fig:0chiral}
\end{figure*}

For the analysis of the superconducting state, the Hamiltonian in Eq.~\ref{eq:H} is decoupled within the Hartree-Fock approximation as
\begin{eqnarray*}
V^{\sigma \sigma{^{'}}} n_{\mathbf{i}\sigma} n_{\mathbf{j}\sigma{^{'}}} \simeq && V^{\sigma \sigma{^{'}}} (\Delta^{\sigma\sigma{^{'}}}_{\mathbf{i}\mathbf{j}} c^{\dagger}_{\mathbf{j} \sigma} c^{\dagger}_{\mathbf{i} \sigma{^{'}}} \\
&+&\bar{\Delta}^{\sigma \sigma{^{'}}}_{\mathbf{i}\mathbf{j}}c_{\mathbf{i}\,\sigma{^{'}}} c_{\mathbf{j}\,\sigma}-|\Delta^{\sigma \sigma{^{'}}}_{\mathbf{i}\mathbf{j}}|^2), \\ \nonumber
U_0 n_{\mathbf{i} \uparrow} n_{\mathbf{i}\downarrow} \simeq && U_0 (\Delta_{0, \mathbf{i}} c^{\dagger}_{\mathbf{i} \uparrow} c^{\dagger}_{\mathbf{i} \downarrow} \\
&+&\bar{\Delta}_{0, \mathbf{i}} c_{\mathbf{i}\downarrow} c_{\mathbf{i}\uparrow}-|\Delta_{0, \mathbf{i}}|^2) .
\end{eqnarray*}
\noindent where the general pairing amplitude on a bond between spin $\sigma$ and $\sigma'$ electrons at the sites ${\mathbf{i}}$ and ${\mathbf{j}}$ is given by $\Delta_{\mathbf{i}\mathbf{j}}^{\sigma \sigma{^{'}}}=\langle c_{\mathbf{i}\,\sigma} c_{\mathbf{j}\,\sigma{^{'}}} \rangle$ and the local singlet is ${\Delta_0}_\mathbf{i}=\langle
c_{\mathbf{i}\downarrow} c_{\mathbf{i}\uparrow} \rangle$.
The numerical analysis consists in evaluating self-consistently these pair correlation amplitudes and, for the $S_z=0$ sector, to combine them to yield the spin-singlet and triplet components as $\Delta^{S,T}_{\mathbf{i}\mathbf{j}}=(\Delta_{\mathbf{i}\mathbf{j}}^{\uparrow \downarrow}\pm\Delta_{\mathbf{j}\mathbf{i}}^{\uparrow \downarrow})/2$.
The solution is obtained by solving the Bogoliubov-de Gennes equations related to the Hamiltonian in Eq.~\ref{eq:H} that gives the spin-resolved energy spectrum of the system, including both bulk and edge Andreev states.

Spin-triplet order parameters can be expressed in a matrix form as~\cite{SigUeda}
\begin{eqnarray}
\Delta_T=
\left(\begin{array}{cc}
  \Delta_{\uparrow\uparrow} & \Delta_{\uparrow\downarrow}\\
  \Delta_{\downarrow\uparrow} & \Delta_{\downarrow\downarrow}
\end{array}\right)
= \left(\begin{array}{cc}
  -d_x+i d_y & d_z \\
  d_z & d_x+id_y
\end{array}\right) \, ,
\label{DeltaT}
\end{eqnarray}
\noindent where the $\vec{d}$-vector components are related to the
pair correlations for the various spin-triplet configurations having zero spin projection
along the corresponding symmetry axis. The three components
$d_x=\frac{1}{2}(-\Delta_{\uparrow\uparrow}+\Delta_{\downarrow\downarrow})$,
$d_y=\frac{1}{2 i}(\Delta_{\uparrow\uparrow}+\Delta_{\downarrow\downarrow})$
and $d_z=\Delta_{\uparrow\downarrow}$ are expressed in terms of
the equal spin $\Delta_{\uparrow\uparrow} ~\mathrm{and}~
\Delta_{\downarrow\downarrow}$, and the anti-aligned spin
$\Delta_{\uparrow\downarrow}$ pair potentials.

For the present study, the pairing interaction $V$ is assumed to be non zero in the
$\uparrow\downarrow$ channel for the chiral case, and in the $\uparrow\uparrow$ and $\downarrow\downarrow$ channels for the helical one.
This implies that the $\vec{d}$-vector is along $z$ for the chiral superconductor, and it lies in the $xy$-plane, which is
chosen to be coincident with the $xy$-plane of the heterostructure, as indicated in Fig.~\ref{fig:1}. Importantly, near the interface, due to the inversion symmetry breaking along the $x$-direction, the triplet order parameter gets mixed with the singlet component within the $S_z=0$ channel.
In the following, we will consider two distinct choices of the triplet vector $\vec{d}_{\bm p}$, both of the chiral type (time reversal symmetry breaking): a) $\vec{d} \equiv (0, 0, p_x+i p_y)$, i.e. $\vec{d}_{\bm p}$ in the $z$-direction, and b) the helical type (time reversal invariant) with $\vec{d} \equiv (p_y, p_x, 0)$, i.e. $\vec{d}_{\bm p}$ lies in the plane of the junction (see Fig. \ref{fig:1}).
Moreover, in order to investigate the effects of the phase difference between the two superconductors, we follow the conventional procedure employed for the study of the Josephson junctions,
by transforming the pairing wave-function in the spin-singlet sides of the heterostructure by the phase factors $\exp[-i \phi/2]$ and $\exp[i \phi/2]$, respectively. By doing so we assume that the domain wall is pinned, and its structure is fixed. In principle supercurrent flowing across the junction may modify the phase profile across the DW, but we leave this extension to future work.

\begin{figure*}[t!]
\includegraphics[width=0.95\textwidth]{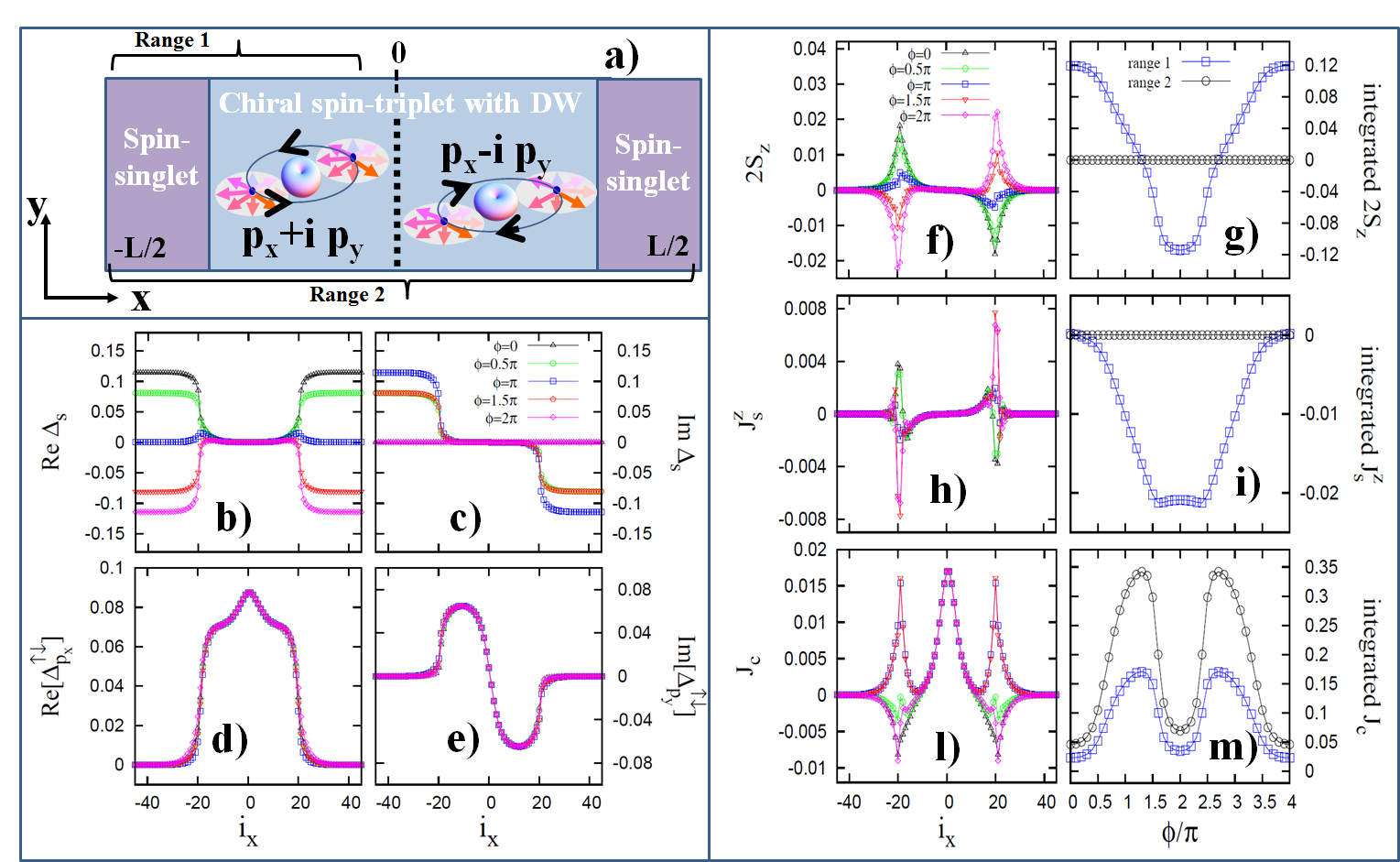}
\caption{(color online). (a) schematic description of the singlet-triplet-singlet (S-T-S) heterostructure with two chiral domains with opposite orbital winding in the spin-triplet region for a spin configuration with zero spin-projection along the $z$ direction (i.e. $d_z$ order parameter). Range 1 (2) indicates half (entire) extension of the S-T-S system. Spatial profile of the real and imaginary parts of the spin-singlet ((b) and (c)) and spin-triplet ((d) and (e)) order parameters is reported at different values of the phase difference $\phi$ between the spin-singlet sides of the heterostructure. (f), (h) and (l) indicate the spatial  evolution along $x$ direction of the $z$-projected magnetization, spin-current and charge-current, respectively. (g), (i) and (m) describe the phase dependent behavior of the integrated quantities over the range 1 (blue squares) and 2 (black circles) for the $z$-projected magnetization, spin-current and charge-current, respectively.}
\label{fig:1chiral}
\end{figure*}

\section{S-T-S heterostructure: chiral and helical domain walls}

As discussed above, interface static magnetic moment, spin, and charge currents can all exist at the boundaries between singlet and triplet superconductors. The question we address below is whether in the S-T-S junction geometry there are measurable differences between the configurations with the single domain triplet and with a domain wall. To this end we present the results comparing the spatial dependence of the self-consistently determined order parameter, magnetization, relevant components of the spin and charge currents for the S-T-S junctions with and without the domain walls for both helical and chiral superconducting triplet order parameter. Moreover, in order to connect our results with those for the case of single interfaces, we always present a comparison of the magnetization or the current integrated over the entire system (two boundaries with/without domain wall) and those integrated over half of the system (a single interface).  The results shown below are for S-T-S junction with size $L=120$ and layers of equal width. Greater values
of $L$ and variation of the pairing coupling amplitudes leave the results
qualitatively unchanged.

\subsection{S-T-S with chiral spin-triplet superconductor}

We start by considering the triplet superconductor with a chiral order parameter. 
We assume that the $\bm d$-vector is fixed by the spin-orbit interaction along the $z$-direction, so that the Copper pairs comprise electron with opposite spins and have total $S_z=0$, see Eq.~\ref{DeltaT}. This phase is stabilized with the choice $V_{\uparrow\downarrow}=2.5$, $V_{\uparrow\uparrow}=V_{\downarrow\downarrow}=0$ and the chemical potential $\mu=-1.8$ (all in units of $t$), which corresponds to filling $n\approx0.4$~\cite{Cuoco08,Romano2010,Kuboki01}.

Fig.~\ref{fig:0chiral} shows the results for the S-T-S system without the domain wall, as sketched in panel a). The real and imaginary components of the singlet and triplet order parameters are shown in panels b), c) and d), e) respectively. As expected, they reach maximal values in the bulk of the corresponding regions. However, 
for the model in Eq. \ref{eq:H} a non-vanishing value of $V_{\uparrow\downarrow}$ promotes coupling both in the singlet and in the triplet channel. In the absence of local inversion symmetry (e.g. near the singlet-triplet interface), the lowest-energy stable configuration has a mixed-symmetry order parameter near the boundary, with non-vanishing spin-triplet and singlet components \cite{Romano2013}. Then, in addition to the proximity effect, the singlet pairing amplitude (Figs.~\ref{fig:0chiral} (b),(c)) is sustained in the T region of the heterostructure close to the interface by that pairing interaction. This mixed parity order parameter is the key player driving the magnetic effects at the singlet-triplet boundary by spin-polarizing the Andreev interface states. Indeed, we observe that a spontaneous magnetization along the $z$-axis develops at the S-T boundary already in the absence of a phase difference, $\phi$ (Figs.~\ref{fig:0chiral} (f)-(g)).
Since the sign of the magnetization is linked to the gradient of the superconducting order parameter near the boundary~\cite{Romano2013}, the ``left'' and the ''right'' S-T boundaries exhibit opposite magnetizations, see Fig.~\ref{fig:0chiral} (f). 
While the magnetization across each interface can be tuned by varying the phase difference, the contributions from the two boundaries remain opposite at any $\phi$, as demonstrated by the vanishing of the total magnetization integrated over the whole system (Fig.~\ref{fig:0chiral} (g)). Strictly speaking, the net magnetization is identically zero only for an ideal symmetric trilayer, but it is reasonable to expect that for asymmetric systems the total magnetization of the heterostructure still undergoes a nearly complete cancellation due to the antiparallel magnetic moments at the two S-T edges. Note that the magnetic moment at a single interface, Fig.~\ref{fig:0chiral} (g), is controlled by the phase difference of $\phi/2$ between the singlet and triplet order parameters, so that the full period of the magnetization dependence on the phase is $4 \pi$.

\begin{figure*}[t!]
\includegraphics[width=0.95\textwidth]{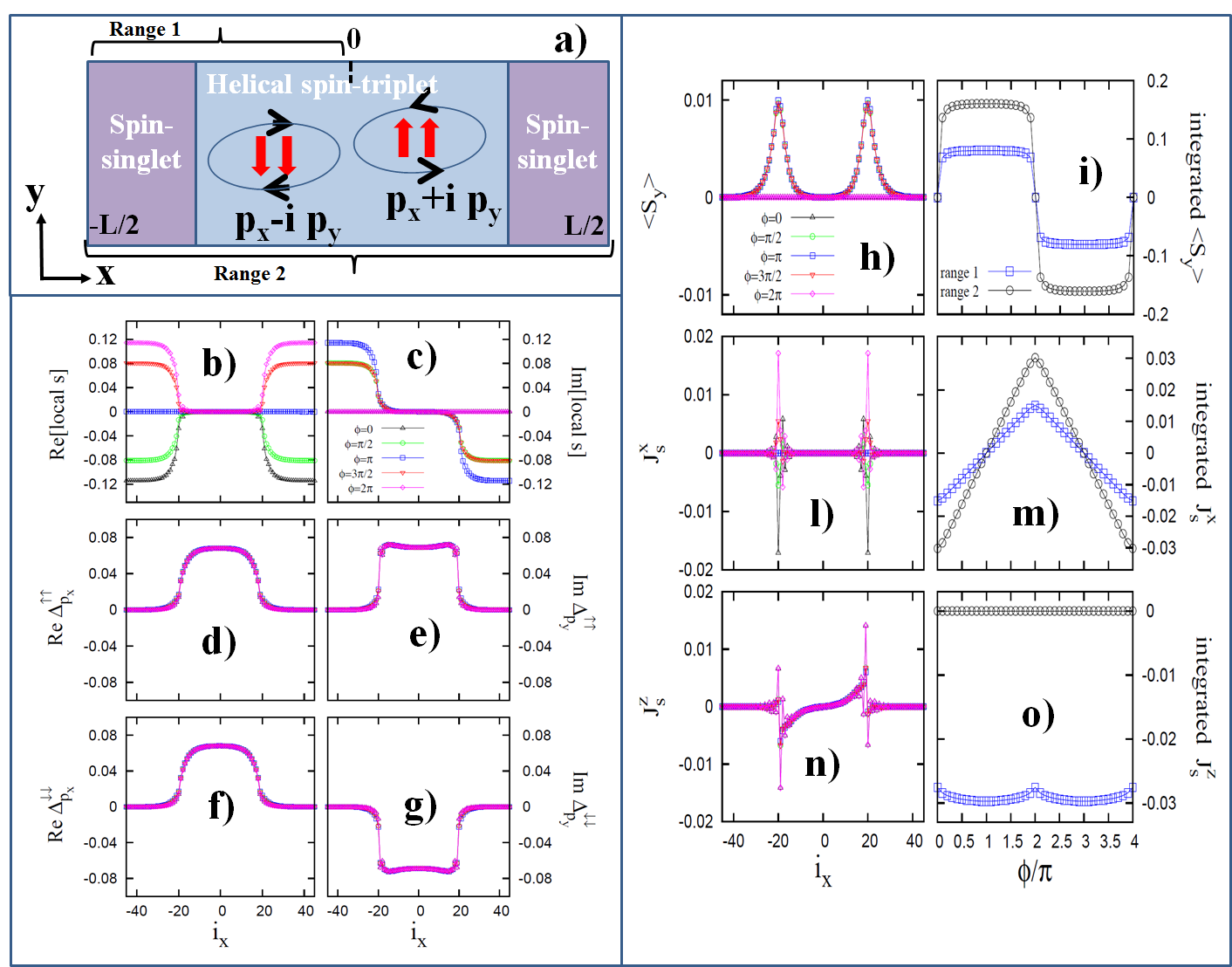}
\caption{(color online). (a) schematic description of the singlet-triplet-singlet (S-T-S) heterostructure with a single helical domain in the spin-triplet region having $p_x+i p_y$ $(p_x-i p_y)$ orbital symmetry for the spin up (down) electron pairs. Range 1 (2) indicates half (entire) extension of the S-T-S system. Spatial profile of the real and imaginary parts of the spin-singlet ((b) and (c)) and spin-triplet ((d)-(e) for up spin and (f)-(g) for down spin polarization) order parameters is shown at different values of the phase difference $\phi$ between the spin-singlet sides of the heterostructure. (h), (l) and (n) indicate the spatial evolution along $x$ direction of the $z$-projected magnetization, $x$- and $z$-projected spin-currents, respectively. (i), (m) and (o) describe the phase dependent behavior of the integrated quantities over the range 1 (blue squares) and 2 (black circles) for the $z$-projected magnetization, $x$- and $z$-projected spin-currents, respectively.}
\label{fig:0helical}
\end{figure*}

However, since the spin-splitting of the Andreev states is opposite at the two interfaces, we find that the net spin current is the same at both, as is seen from Figs.~\ref{fig:0chiral} (h),(i), and adds up to a non-zero net $J_s^z$, whose amplitude can be modulated by the phase difference. As expected for the chiral spin-triplet superconductor a finite charge current $J_{c}$ flows in opposite direction at each S-T interface, see Figs.~\ref{fig:0chiral} (l),(m). This current is carried by the Andreev bound states at each interface, and therefore the same spin-splitting of these states that leads to the finite interface magnetization reduces the magnitude of $J_c$.
Consequently, the maximum of the charge current at each interface is reached not at $\phi=0$, but for the phase difference with vanishing magnetization, at $\phi \sim 1.25 \pi$ in our case. Quite generally, changing the phase difference across the interface has a complex effect on the dispersion of the bound states, which, at a finite $\phi$, is determined not only by the mismatch between the singlet and the triplet order parameters, but also by the phase difference between the $s$-wave component of the mixed parity state on the triplet side, and the corresponding isotropic order parameter of the $s$-wave lead. As a result, the relation between the spin and charge current amplitudes at $\phi=0$ and $\phi=2\pi$ ($\pi$ phase shift across a single interface) is non-trivial.

We foresee the possibility to coherently switch the system from a state that has only charge current close to the edge (e.g. at $\phi=0$) to another one where both spin and charge currents are present at each single interface (e.g. at $\phi=\pi$). Note that once we sum over the contributions of the two interfaces, due to the chirality of the order parameter, we have only a net spin current whose amplitude can be phase modulated, see Figs.~\ref{fig:0chiral} (h),(i).

We now compare these results with the behavior of the equivalent junction where a triplet layer contains two domains with opposite chirality, i.e. with an orbital content of the form $p_x + ip_y$ and $p_x-ip_y$, respectively, see Fig.~\ref{fig:1chiral} (a). We consider the geometry with the domain wall parallel to the interface.
The evolution of the order parameters (Figs. \ref{fig:1chiral} (b)-(e)) is akin to that obtained for the case without a domain wall except that the imaginary part of the triplet $p_y$ component of the triplet changes sign across the domain wall at $i_x=0$. Concomitantly, this leads to a moderate enhancement of the $p_x$ component at the DW location. Since the interface magnetization occurs only at the boundary with the singlet supporting the mixed parity order parameter, there is no change in its behavior between the two geometries, compare Figs.~\ref{fig:1chiral} (f)-(g) with the corresponding panels of Fig. \ref{fig:0chiral}. This identical behavior also supports our understanding that the origin of the magnetization is in the coupling of the gradient of the $p_x$ component of the triplet order parameter to the singlet superconductivity~\cite{Romano2013,Romano2016}. This component turns out to be unchanged across the domain wall. On the other hand, for the chiral superconductors the dispersion of the surface state is determined by the relative phase of the $p_x$ and $p_y$ components as well as the ``left'' or ``right'' orientation of the boundary, and therefore now the velocity of the Andreev states is the same at the two S-T interfaces. Consequently the contributions to the spin current from the two interfaces are opposite,  Fig.~\ref{fig:1chiral}(h), and the net spin current vanishes, Fig.~\ref{fig:1chiral}(i), along with the net magnetization.

On the other hand, as schematically depicted in Fig. 1 and shown in Fig.~\ref{fig:1chiral}(l), for the same reason now the charge currents flow in the same direction at both S-T interfaces. At the same time along the line $i_x=0$ there is an additional charge current due to the opposite orbital circulation of the Cooper pairs in the two domains. While the former contribution can be controlled by the phase difference across the heterostructure, the latter is phase-insensitive. Consequently, while the overall shape of the phase dependence of $J_c$ in Fig.~\ref{fig:1chiral}(m) is similar to that found for half-junction in in Fig.~\ref{fig:0chiral}(m), there is an overall shift due to the domain wall, so that the sign of the net current does not change. 


The main conclusion from this comparison is that in an S-T-S heterostructure with chiral spin-triplet and one component of the $\bm{d}_{\bm{k}}$-vector,  perpendicular to the S-T-S planar junction in our case, the presence of a domain wall allows separate control of the net charge and spin currents flowing through the spin-triplet region. For the  single-domain configuration the phase difference across the junction tunes a non vanishing spin current in the absence of a charge current, whereas in the case of double chiral domains the phase difference can drive a net charge current in the absence of a net spin flow. Such observation can be immediately extended to the general case of even and odd number of chiral domain walls within the spin-triplet superconductor. Even (odd) number of chiral domain walls would then yield a modulated net spin (charge) current flow across the spin-triplet superconducting region.

\subsection{S-T-S with helical spin-triplet superconductor}

By choosing the pairing coupling such as $V_{\uparrow\uparrow}=V_{\downarrow\downarrow}\ne0$ and $V_{\uparrow\downarrow}=0$ the superconducting region can exhibit a stable spin-triplet state with helical $\vec{d}=(p_y,p_x,0)$ symmetry. The superconducting pairing now occurs in the equal spin channels, $\vec{d}=\widehat{\bm x} p_y+\widehat{\bm y} p_x$, and therefore it is natural to contrast this case with the previously considered chiral opposite spin pairing. Moreover, there are suggestions~\cite{Scaffidi} that the helical order is close in energy to the chiral paired state in Sr$_2$RuO$_4$, making such a comparison necessary in order to help in determining the order parameter for this candidate triplet superconductor.

\begin{figure*}[t][h]
\includegraphics[width=0.95\textwidth]{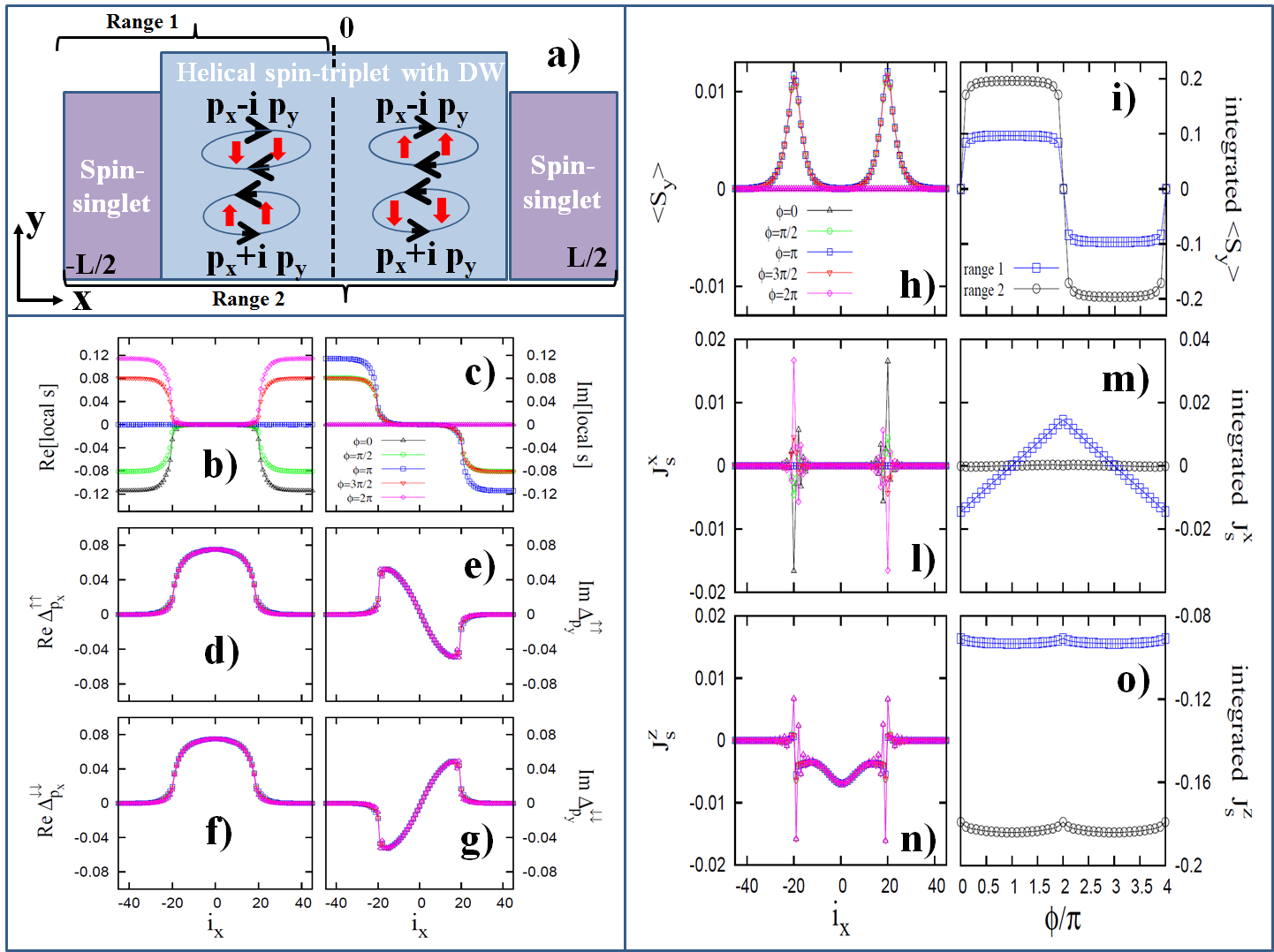}
\caption{(color online). (a) schematic description of the singlet-triplet-singlet (S-T-S) heterostructure with two helical domains. Range 1 (2) indicates half (entire) extension of the S-T-S system. Spatial profile of the real and imaginary parts of the spin-singlet ((b) and (c)) and spin-triplet ((d)-(e) for up spin and (f)-(g) for down spin) order parameters is shown at different values of the phase difference $\phi$ between the spin-singlet sides of the heterostructure. (h), (l) and (n) indicate the spatial evolution along the $x$ direction of the $z$-projected magnetization, $x$- and $z$-projected spin-currents, respectively. (i), (m) and (o) describe the phase dependent behavior of the integrated quantities over the range 1 (blue squares) and 2 (black circles) for the $z$-projected magnetization, $x$- and $z$-projected spin-currents, respectively.}
\label{fig:1helical}
\end{figure*}

As illustrated in Fig.~\ref{fig:0helical}(a), the pairing state can be thought of as a superposition of the $p_x\pm ip_y$ states for the spin-up and spin-down Copper pairs, respectively. The self-consistently determined order parameters are shown in Fig.~\ref{fig:0helical}(b)-(g). The essential difference now is that there is no direct coupling between the gradient of any equal-spin triplet component, the singlet order parameter, and the magnetization, and therefore no static spin polarization appears at the interface in the absence of a phase difference between the S layers. However, as one can see from Fig.~\ref{fig:0helical} (h),
finite $\phi$ gives rise to a spin polarization in the $y$ direction which, in a sharp contrast to the chiral case,
has the same sign at the two S-T interfaces. The origin of this net magnetic moment is once again the spin splitting of the Andreev bound states at the interface, which, in the absence of the mixed symmetry order parameters, is in exact analogy to the situation studied in Ref.~\onlinecite{Yakovenko}

It is well known that both in semiconductors~\cite{Mishchenko} and in superconductors with antisymmetric spin-orbit coupling~\cite{Vorontsov}, near the interface the spins deviate from the principal quantization axis in the bulk, resulting in the spin current of the component normal to that axis. In our case similar physics arises from the need to rotate from the equal-spin pairing amplitude in the triplet phase to the opposite spin pairing on the singlet side: there is an effective spin-active interface leading to the appearance of the spin current of the transverse component. Mirror symmetry of the system prevents the appearance of the spin current polarized along the $y$ direction, and hence we find solely a current of carriers with spins polarized in the $x$ direction, Fig.~\ref{fig:0helical} (l).  

Finally, the counterclockwise rotating spin-up and clockwise rotating spin-down Cooper pairs naturally give rise to the edge spin current of the $z$-component in spin space, Fig.~\ref{fig:0helical}(n). As is clear from Fig.~\ref{fig:0helical}(a), the net $J_s^z$ is directed down (up) on the left (right) S-T interface, and therefore averages to zero over the entire junction, Fig.~\ref{fig:0helical}(o). Note that both spin currents exhibit spatial oscillations near the interface on the scale roughly equal to the coherence length, Fig.~\ref{fig:0helical}(l),(n).

As before, splitting the helical superconductor into two domains does not affect the behavior at the interface, and the net magnetic moment, Fig.~\ref{fig:1helical}(h), (i), which still comes from the parity mixing at the singlet-triplet interfaces. The same interfaces give dominant contribution to the spin current of the in-plane component. Therefore, with the domain wall, these currents flow in the opposite directions over the two S-T boundaries, and average to zero irrespective of the phase difference across the junction.  With the two domains having opposite circulation of the Cooper pairs with each spin polarization, the spin currents of the $z$ component add along the domain wall, yielding a large contribution, Fig.~\ref{fig:1helical}(n). The S-T interfaces support $J_s^z$ in the direction opposite to the that at the domain wall, but, because of the suppression of the superconducting order parameter, those are smaller. The net values of this component therefore remains finite when integrated across the junction, Fig.~\ref{fig:1helical}(o), and has a greater magnitude than the corresponding value for the system without the domain wall, as compared with Fig.~\ref{fig:0helical}(o).

\section{Conclusions}

We have determined the behavior of transport and magnetic properties of an S-T-S heterostructure by considering single and double-domain structure of the spin-triplet region for both chiral and  helical order parameters, and investigated the dependence of these properties on the phase difference across the junction. The static magnetization in all situations is due to the parity mixing at the singlet-triplet interface, and is confined to the boundary layers, and therefore is insensitive to the existence of the domain wall. For chiral superconductors with $S_z=0$ opposite spin-triplet pairing, the interface magnetization normal to the plane appears at the phase difference $\phi=0$, while for the helical pairing the magnetization is parallel to the interface direction, and only exists for a finite phase difference across the heterostructure. 

The charge current only appears for the chiral order parameter, and does not average to zero if two different domains are present due to the large contribution of the domain wall. Importantly, the phase dependence of this current is complex due to the splitting of the interface states by the static magnetization, and reaches maximal value precisely when the interface magnetization vanishes, at the value of the phase that is determined by the details of the pairing interaction.

Similarly, the spin current of the in-plane spin component only appears for the helical case due to the mismatch of the spin structures of the Cooper pairs in the singlet and triplet regions, it is always confined to the S-T boundaries, but only averages to zero in the presence of a domain wall. In the single domain case this component of the spin current disappears for $\phi=\pi$.

The spin current of the out-of-plane spin component exists in all the considered cases, but averages to zero for single domain helical and double domain chiral triplet states. For the double domain helical triplet the dominant contribution to $J_s^z$ is from the domain wall region, in analogy to the charge current in the chiral case. 

In summary, considering the domain walls in the triplet superconductors as parts of an S-T-S junction considerably expands the range of control of spin and charge currents, and offers new pathways towards identification of non-trivial superconducting orders.

\section*{Acknowledgments}
This research received support from NSF DMR Grant No 1410741 (I. V.).

\end{document}